\documentclass[a4paper,fleqn,usenatbib]{mnras}

\usepackage{newtxtext,newtxmath}

\usepackage[T1]{fontenc}
\usepackage{ae,aecompl}
\usepackage{threeparttable}

\usepackage{graphicx}	
\usepackage{amsmath}	
\usepackage{booktabs,caption}	
\usepackage{multicol}
\DeclareMathOperator{\sech}{sech}

\title [Vertical distribution in a non-isothermal disc]{Vertical stellar density distribution in a non-isothermal galactic disc}

\author[S. Sarkar and C.J. Jog ]
       {Suchira Sarkar$^{1}$\thanks{E-mail : suchira@iisc.ac.in}, and
        Chanda J. Jog$^{1}$\thanks{E-mail : cjjog@iisc.ac.in}\\
$^1$   Department of Physics,
Indian Institute of Science, Bangalore 560012, India \\
} 

\date{Accepted XXX. Received YYY; in original form ZZZ}

\pubyear{2020}

\begin{document}
\label{firstpage}
\pagerange{\pageref{firstpage}--\pageref{lastpage}}
\maketitle

\begin{abstract}

The vertical density distribution of stars in a galactic disc is traditionally obtained by assuming an isothermal vertical velocity dispersion of stars. Recent observations from SDSS, LAMOST, RAVE, \textit{Gaia} etc show that this dispersion increases with height from the mid-plane. Here we study the dynamical effect of such non-isothermal dispersion on the self-consistent vertical density distribution for the thin disc stars in the Galaxy, obtained by solving together the Poisson equation and the equation of hydrostatic equilibrium. We find that in the non-isothermal case the mid-plane density is lower, and the scale height is higher than the corresponding values for the isothermal distribution, due to higher vertical pressure, hence the distribution is vertically more extended. The change is  $\sim 35 \%$ at the solar radius for a stars-alone disc for the typical observed linear gradient of $+6.7~\mathrm{km~s^{-1}kpc^{-1}}$ and becomes even higher with increasing radii and increasing gradients explored. The distribution shows a wing at high $z$, in agreement with observations, and is fitted well by a double $\sech^{2}$, which could be mis-interpreted as the existence of a second, thicker disc, specially in external galaxies. We also consider a more realistic disc consisting of gravitationally coupled stars and gas in the field of dark matter halo. The results show the same trend but the effect of non-isothermal dispersion is reduced due to the opposite, constraining effect of the gas and halo gravity. Further, the non-isothermal dispersion lowers the theoretical estimate of the total mid-plane density i.e, Oort limit value, by 16\%.

\end{abstract}

\section{Introduction}

The vertical density distribution of stars in a galactic disc is an important property in the study of disc structure and dynamics. The vertical distribution can be obtained in a self-consistent way by solving the Poisson equation and the vertical Jeans equation (or equation of hydrostatic equilibrium) together, as was first shown in the classic work by Spitzer (1942). This was followed up by several studies (e.g. Bahcall 1984a,b,c; Bahcall, Flynn \& Gould 1992). The vertical structure of a coupled multi-component disc of stars plus gas in a dark matter halo potential has also been studied by various authors (Narayan \& Jog 2002; Kalberla 2003; Banerjee \& Jog 2007; Comer$\mathrm{\acute{o}}$n et al. 2011; Sarkar \& Jog 2018).

Note that these assume an isothermal (or a constant) stellar velocity dispersion along vertical direction for simplicity. In fact, traditionally, most papers in the literature on the galactic vertical structure make this assumption, so that the general, non-isothermal case is rarely discussed.

Only a few papers have examined the consequences of considering a non-isothermal velocity dispersion (Camm 1950; Perry 1969).
The reason could be that the velocity dispersion was not known in detail at different heights. Interestingly, Camm (1950) and Perry (1969) had studied the vertical distribution for the case when the velocity dispersion increases with height from the mid-plane and found that a stable solution was possible. 

In the context of the study of the Oort limit, it was realized that a non-isothermal dispersion is needed to satisfy the observational constraints.
A superposition of separate isothermal distribution for tracers was used to mimic this (Oort 1960; Bahcall 1984a,b,c; Kuijken \& Gilmore 1989).
 
The data were not available to enable one to study the deviation from isothermal behavior in well defined, homogeneously selected samples of stars(Bahcall 1984a). Thus these studies did not consider a variation of dispersion with height as a physical feature and they did not obtain the self-consistent stellar distribution as done in the present paper.
Thus the non-isothermal vertical stellar disc modelling remained of academic interest only.

This has changed in recent years for two reasons. First, the numerical simulations show that a tidal encounter can heat up the disc so that the stellar vertical dispersion increases with time (Walker, Mihos \& Hernquist 1996;  Vel$\mathrm{\acute{a}}$zquez \& White  1999), although the vertical variation of the velocity dispersion has not been studied explicitly in these simulations. While the topic of dynamical, secular heating of stars has been studied extensively, these studies show that the heating is more effective for the planar dispersion. The typical physical processes considered are star-cloud encounters (Lacey 1984) and star-spiral arm encounters (Barbanis \& Woltjer 1967; Carlberg \& Sellwood 1985). Later studies show that the vertical dispersion could also increase with time due to the heating by spiral arms (Jenkins \& Binney 1990) and also bars (Saha, Tseng \& Taam 2010). These studies were motivated by, and explain fairly well, the observed increase in stellar velocity dispersion with age (Wielen 1977). However, these theoretical studies have not studied the possible vertical variation of the stellar velocity dispersion or how the gradient varies with time. Recent simulations too focus on the time evolution of the dispersions due to these secular heating processes (Aumer et al. 2016; Gustafsson et al. 2016; Wu et al. 2020). We note that if the variation is due to tidal interactions then the velocity dispersion would be expected to increase with distance from the mid-plane at low distances while a falling dispersion would be expected if the variation is due to energy input due to supernovae.

Second, with the advent of high-quality, high-resolution data, such as obtained using Sloan Digital Sky Survey (SDSS), Large sky area multi-object fiber spectroscopic telescope (LAMOST), Radial Velocity Experiment (RAVE), \textit{Gaia} etc, a variation of stellar velocity dispersion with height for various tracers has now been observed, even over different metallicity bins, and discussed in the solar neighbourhood (Fuchs et al. 2009; Bienaym$\mathrm{\acute{e}}$ et al. 2014; Binney et al. 2014; Jing et al. 2016; Xia et al. 2016; Hagen \& Helmi 2018; Gaia collaboration et al., 2018; Guo et al.,2020; Salomon et al.,2020) and also at larger radii (Bond et al. 2010; Sharma et al. 2020; Sun et al. 2020). This shows that the non-isothermal velocity dispersion is a genuine physical feature of the stellar disc. This has been our motivation to look at this problem. A few of the above papers discuss the vertical variation of the dispersion, particularly for the thin disc of stars (Jing et al. 2016; Xia et al. 2016; Hagen \& Helmi 2018; Guo et al. 2020; Sun et al. 2020).
Although a few papers, based on the observed data, have explicitly discussed the effect of non-isothermal dispersion on the measurement of dynamical quantities, such as local dark matter estimate (Garbari, Read \& Lake 2011), the effect on the vertical density distribution and hence the shape of the density profile has not been studied so far.

Here we study the dynamical effect of a non-isothermal vertical velocity dispersion on the vertical density distribution for the Galactic thin disc stars. In order to isolate this effect we first consider a stars-alone disc, for which the isothermal distribution gives the well-known density distribution of $\sech^{2}$ form (Spitzer 1942). We then examine the consequences for the more realistic case of a multi-component, coupled stars plus gas disc in the gravitational field of a dark matter halo. We find that the inclusion of non-isothermal dispersion has a significant effect on the vertical distribution. 

We note that we model the stellar disc to consist of a single component, for simplicity, i.e., we have treated stars of different ages or metallicity values of the thin disc cumulatively. This is a common practice in studies of vertical disc structure of galaxies (e.g., Spitzer 1942). For example, even in the determination of the stellar velocity dispersion (eq. 16), the dispersion measured is used to denote the average value. 

We discuss the formulation of the model equations and the observed vertical velocity dispersion gradients as well as other input parameters used in our model, in Section 2. Section 3 contains the results for the non-isothermal vertical distribution of stars for both stars-alone case and a realistic, gravitationally coupled multi-component system of stars, gas and dark matter halo. Section 4 contains the implication of non-isothermal dispersion on the determination of the Oort limit. Sections 5 contains a brief discussions on the limit of validity of isothermal assumption. Finally we present the conclusions in Section 6.

We note that it is interesting to study if the non-isothermal vertical velocity dispersion can affect the calculation of important kinematical quantities. One such parameter is the asymmetric drift, defined for a stellar population, which explicitly depends on the velocity dispersions. We discuss the possible effects of non-isothermal dispersion in this case briefly in the Appendix A.

\section{Formulation of the Problem}

\subsection{Stars-alone, and multi-component disc models}

We use galactocentric cylindrical coordinates ($R$, $\phi$, $z$). The vertical hydrostatic equilibrium (balance) equation for an axisymmetric stars-alone disc is given as (Rohlfs 1977)

\begin{equation}
\frac{1}{\rho}\frac{\mathrm{d}(\rho\sigma^{2}_{z})}{\mathrm{d}z} = K_{z}\label{eq:1}
\end{equation}

\noindent where $\rho$ is mass density, $\sigma_{z}( =\langle(v_{z})^2\rangle^{1/2})$ is the vertical velocity dispersion of stars and $v_{z}$ is the vertical velocity. $K_{z}( = -\mathrm{d}\Phi/\mathrm{d}z)$ denotes the vertical force per unit mass due to stars where $\Phi$ is the corresponding potential.

\noindent The Poisson equation for a thin stars-alone disc is 
\begin{equation}
\frac{\mathrm{d}^{2}\Phi}{\mathrm{d}z^{2}} = 4 \pi G \rho. \label{eq:2}
\end{equation}

\noindent These two equations can be combined together to write the following joint hydrostatic balance-Poisson equation as
\begin{equation}
\frac{\mathrm{d}}{\mathrm{d}z}\left[\frac{1}{\rho}\frac{\mathrm{d}(\rho \sigma^{2}_{z})}{\mathrm{d}z}\right] = -4 \pi G \rho .\label{eq:3}			
\end{equation}

\noindent When the dispersion is isothermal, i.e, constant along $z$, the above equation reduces to 

\begin{equation}
\sigma^{2}_{z}\frac{\mathrm{d}}{\mathrm{d}z}\left[\frac{1}{\rho}\frac{\mathrm{d}\rho}{\mathrm{d}z}\right] = -4 \pi G \rho. \label{eq:4}	
\end{equation}. 

\noindent The solution is obtained as (Spitzer 1942)
\begin{equation}
\rho(z)=\rho_{0}\sech^{2}(z/z_{0})\label{eq:5}
\end{equation}
\noindent referred to as the $\sech^{2}$ model describing the vertical distribution of stars for a stars-alone isothermal disc. Here $\rho_{0}$ denotes the mid-plane density and $z_{0}$ is a measure of the scale height. Now when the vertical velocity dispersion becomes a function of $z$, Eq.(3) can be written as

\begin{equation}
\frac{\sigma^{2}_{z}}{\rho}\frac{\mathrm{d}^{2}\rho}{\mathrm{d}z^{2}}-\frac{\sigma^{2}_{z}}{\rho^{2}}\left(\frac{\mathrm{d}\rho}{\mathrm{d}z}\right)^{2}+\frac{1}{\rho}\left(\frac{\mathrm{d}\rho}{\mathrm{d}z}\right)\frac{\mathrm{d}\sigma^{2}_{z}}{\mathrm{d}z}+\frac{\mathrm{d}^{2}\sigma^{2}_{z}}{\mathrm{d}z^{2}}  = -4 \pi G \rho. \label{eq:6}
\end{equation}

\noindent 
Any general analytic form of velocity dispersion can be used in this equation. In this paper, we consider the vertical velocity dispersion $\sigma_{z}$ to increase linearly along $z$, based on most of the observed data in literature (discussed later in Sec 2.2.1), and write its expression as

\begin{equation}
\sigma_{z}=\sigma_{z,0}+Cz \label{eq:7}
\end{equation} 

\noindent where "$C$" is the linear gradient ($\mathrm{d}\sigma_{z}/\mathrm{d}z$) in velocity dispersion along $z$ and $\sigma_{z,0}$ is the dispersion at $z$=0, i.e, the galactic mid-plane. We substitute this expression into Eq.(6) and obtain 

\begin{equation}
\begin{split}
\frac{\mathrm{d}^{2}\rho}{\mathrm{d}z^{2}} & = \frac{-4\pi G \rho^{2}}{\left(\sigma_{z,0}+Cz\right)^{2}}+\frac{1}{\rho}\left(\frac{\mathrm{d}\rho}{\mathrm{d}z}\right)^{2} \\
		         &  -\frac{2C}{\left(\sigma_{z,0}+Cz\right)}\left(\frac{\mathrm{d}\rho}{\mathrm{d}z}\right)-\frac{2C^{2}\rho}{\left(\sigma_{z,0}+Cz\right)^{2}}.\label{eq:8}
\end{split}
\end{equation}

\noindent The solution of this equation gives the self-consistent non-isothermal vertical density distribution $(\rho(z))$ of stars-alone disc for a linearly increasing vertical velocity dispersion. We assume that such a stable solution exists.
We solve Eq.(8) by applying the fourth-order Runge-Kutta method, where the observed surface density of stars is used as one boundary condition and the second condition is given by $\mathrm{d}\rho/\mathrm{d}z=0$, defined at $z$ = 0 (Narayan \& Jog 2002; Sarkar \& Jog 2018). This condition is satisfied for any realistic vertical density distribution which is homogeneous close to the mid-plane. The corresponding isothermal case is solved by setting the gradient $C=0$. Alternately the solution in this case is simply given as a $\sech^{2}$ law (Eq. 5).

Now for a realistic system of gravitationally coupled multi-component disc of stars, and interstellar gas (atomic hydrogen gas, $\mathrm{HI}$ \& molecular hydrogen gas, $\mathrm{H_{2}}$) in the field of dark matter halo, the hydrostatic balance equation for any component is given as

\begin{equation}
\frac{1}{\rho_{i}}\frac{\mathrm{d}(\rho_{i}\sigma^{2}_{z,i})}{\mathrm{d}z} = (K_{z})_{\mathrm{stars}}+(K_{z})_{\mathrm{HI}}+(K_{z})_{\mathrm{H_{2}}}+(K_{z})_{\mathrm{DM}}\label{eq:9}
\end{equation}
\noindent where $i$=stars, HI and $\mathrm{H_{2}}$. The Poisson equation for the thin disc of stars plus gas is

\begin{equation}
\frac{\mathrm{d}^{2}\Phi_{\mathrm{stars}}}{\mathrm{d}z^{2}}+\frac{\mathrm{d}^{2}\Phi_{\mathrm{HI}}}{\mathrm{d}z^{2}}+\frac{\mathrm{d}^{2}\Phi_{\mathrm{H_{2}}}}{\mathrm{d}z^{2}} =4 \pi G(\rho_{\mathrm{stars}}+\rho_{\mathrm{HI}}+\rho_{\mathrm{H_{2}}}).\label{eq:10}
\end{equation}
\noindent Thus using Eq.(9) \& 10, the joint hydrostatic balance - Poisson equation for each of the disc components (i=stars, HI and $\mathrm{H_{2}}$ gas respectively) can be written as 

\begin{equation}
\frac{\mathrm{d}}{\mathrm{d}z}\left[\frac{1}{\rho_{i}}\frac{\mathrm{d}(\rho_{i}\sigma^{2}_{z,i})}{\mathrm{d}z}\right] =-4 \pi G(\rho_{\mathrm{stars}}+\rho_{\mathrm{HI}}+\rho_{\mathrm{H_{2}}})+\frac{\mathrm{d}(K_{z})_{\mathrm{DM}}}{\mathrm{d}z}.\label{eq:11}
\end{equation}
\noindent Putting the expression of Eq.(7) into the above equation, we derive the following equation which governs the non-isothermal vertical distribution of stars in the realistic system,

\begin{equation}
\begin{split}
\frac{\mathrm{d}^{2}\rho_{\mathrm{stars}}}{\mathrm{d}z^{2}} & = \frac{\rho_{\mathrm{stars}}}{\left(\sigma_{z,0}+Cz\right)^{2}}\left[-4\pi G\left(\rho_{\mathrm{stars}}+\rho_{\mathrm{HI}}+\rho_{\mathrm{H_{2}}}\right)+\frac{\mathrm{d}(K_{z})_{\mathrm{DM}}}{\mathrm{d}z}\right]  \\
					& +\frac{1}{\rho_{\mathrm{stars}}}\left(\frac{\mathrm{d}\rho_{\mathrm{stars}}}{\mathrm{d}z}\right)^{2} \\
					& -\frac{2C}{\left(\sigma_{z,0}+Cz\right)}\left(\frac{\mathrm{d}\rho_{\mathrm{stars}}}{\mathrm{d}z}\right)-\frac{2C^{2}\rho_{\mathrm{stars}}}{\left(\sigma_{z,0}+Cz\right)^{2}}. \label{eq:12}
\end{split}
\end{equation}

\noindent We consider HI and $\mathrm{H_{2}}$ gas both to have isothermal dispersion along $z$, as supported by observed data, and therefore the following equation governs their vertical distribution:

\begin{equation}
\begin{split}
\frac{\mathrm{d}^{2}\rho_{i}}{\mathrm{d}z^{2}} & = \frac{\rho_{i}}{\sigma^{2}_{z,i}}\left[-4\pi G\left(\rho_{\mathrm{stars}}+\rho_{\mathrm{HI}}+\rho_{\mathrm{H_{2}}}\right)+\frac{\mathrm{d}(K_{z})_{\mathrm{DM}}}{\mathrm{d}z}\right]  \\
                           &   +\frac{1}{\rho_{i}}\left(\frac{\mathrm{d}\rho_{i}}{\mathrm{d}z}\right)^{2} \label{eq:13}
\end{split}
\end{equation}

\noindent for $i$= HI \& $\mathrm{H_{2}}$ respectively. We solved Eq.(12) \& Eq.(13) together for a  coupled system, following the approach as in Narayan \& Jog (2002). The solutions for stars, HI \& $\mathrm{H_{2}}$ are obtained in an iterative fashion till fifth decimal convergence in the solution of each component, following the same boundary conditions as discussed for the stars-alone disc.

We consider the dark matter halo profile to be pseudo-isothermal, expressed in the spherical polar coordinate as (Mera, Chabrier \& Schaeffer 1998):

\begin{equation}
\rho_{\mathrm{DM}}(r)=\frac{V^{2}_{\mathrm{rot}}}{4\pi G}\frac{1}{\left(R^{2}_{\mathrm{c}}+r^{2}\right)}\label{eq:14}
\end{equation}
\noindent where $R_{\mathrm{c}}$ is the core radius and $V_{\mathrm{rot}}$ is the limiting rotation velocity and $r$ is the radius in the spherical polar coordinates. The potential for this profile is, 

\begin{equation}
\Phi_{\mathrm{DM}}(r)=-V_{\mathrm{rot}}^{2}\left[1-\frac{1}{2}\log\left(R^{2}_{\mathrm{c}}+r^{2}\right)-\frac{R_{\mathrm{c}}}{r}\tan^{-1}\left(\frac{r}{R_{\mathrm{c}}}\right)\right]. \label{eq:15}
\end{equation}

\noindent Using cylindrical coordinates we write $\mathrm{d}(K_{z})_{\mathrm{DM}}/\mathrm{d}z =-\partial^{2}\Phi_{\mathrm{DM}}/\partial z^{2} $ as the halo contribution (as given in Narayan \& Jog (2002), with an additional negative sign multiplying the total expression, since the negative sign was missed as a typographical error in the expression for the potential in that paper).

\subsection{Input parameters}

The above formulation is general for any disc galaxy, we now apply it to the Milky Way since the variation in $\sigma_{z}$ is known for it observationally. The $\sigma_{z}$ vs. $z$ profile for thin disc stars is discussed in the literatures mainly within a few kpc of the solar neighbourhood, or $R\sim$ 6-10 kpc. Therefore we apply the model at $R$=6,8.5 (taken to be the solar radius) and 10 kpc. These choices will help bring out the dynamical effect of non-isothermal dispersion with increasing radii where the self-gravity of the disc becomes successively lower.

\subsubsection{Observed non-isothermal vertical velocity dispersion of stars}

The radial velocity dispersion values of stars at mid-plane were obtained observationally by Lewis \& Freeman (1989), which fall off exponentially with radius as :
\begin{equation}
\sigma_{R}= 105 \exp(-R/8.7~\mathrm{kpc}) ~\mathrm{km~s^{-1}}.\label{eq:16}
\end{equation}

\noindent From these the corresponding vertical velocity dispersion values at the mid-plane, i.e, $\sigma_{z,0}$ were calculated by assuming the vertical to radial dispersion ratio to be 0.45 (Dehnen \& Binney 1998; Mignard 2000), as observed in the solar neighbourhood and assumed to be the same here at all radii. At each radius, we apply the observed gradient value along $z$, which is discussed next.

Hagen \& Helmi (2018) characterized $\sigma_{z}$ for stars in the region $R\sim$ 6-10 kpc using the red clump stars as tracers, by analyzing the observed data from TGAS \& RAVE surveys. The authors mention that they fit a linear function to the dispersion values around the solar radius and thus derive the gradient from the slope of the fitted curve. Although the slope is not mentioned in their paper, we note that the net increase in $\sigma_{z}$ in the thin disc is found to be around $\sim \mathrm{10 ~km~s^{-1}}$ over the observed $z$ interval of 1.5 kpc from the mid-plane (figs 5 \& 6 in their paper). This leads to a linear velocity dispersion gradient value of $\sim$+6.7 $\mathrm{km~s^{-1}kpc^{-1}}$. 

Recently, Guo et al.(2020) discussed the kinematics of a cross-matched sample of G/K-type dwarf stars from LAMOST DR5 and \textit{Gaia} DR2 in the solar neighbourhood. The stars are chosen from the thin disc and contain only a specific spectral type. The vertical velocity dispersion profile, studied upto $|z|$=1.3 kpc, is found to be increasing along $z$ in a similar fashion as seen in Hagen \& Helmi (2018). Earlier, Jing et al.(2016) quoted a linear gradient value of +7.2 $\mathrm{km~s^{-1}kpc^{-1}}$ for the thin disc stars, observed in the range of 6.5 kpc$<R<$9.5 kpc within 0.1 kpc$<|z|<$3 kpc, obtained using F/G-type dwarf stars as tracers from the SDSS \& LAMOST survey. Xia et al.(2016) also showed the increase in $\sigma_{z}$ for the thin disc stars from $z\sim$ 200 to $\sim$ 1500 pc, using K \& G type main-sequence stars from LAMOST. Sun et al. (2020) also showed a similar increase within $|z|$<3 kpc using the thin disc sample of red clump stars from LAMOST \& \textit{Gaia}.

Thus, the observed vertical velocity dispersion profiles of the thin disc stars in the solar neighbourhood are found to be comparable to each other within a small range irrespective of the tracer or survey and therefore the measured linear gradient values are expected to lie within a small range. Thus irrespective of the actual gradient value chosen, the trend in the results will be similar. We also note that, these dispersion profiles are obtained for the thin disc tracer populations that consist of a finite range of metallicities and ages, rather than a single value of metallicity or age. Therefore the resulting gradient value is applicable for the whole thin disc treated as a single component in our work (see Section 1).

Based on these arguments, we choose the gradient value of +6.7 $\mathrm{km~s^{-1}kpc^{-1}}$, (Hagen \& Helmi 2018) as the standard value to explore the effect of non-isothermal velocity dispersion. We assume that the same gradient value ($\mathrm{|6.7|~km~s^{-1}~kpc^{-1}}$) will be applicable for both $|z|$-directions.

We note that, a gradient value higher than the observed ones may be possible as well, specially in the outer disc due to the possibility of external tidal interactions. Therefore, we also consider a higher dispersion gradient of +10  $\mathrm{km~s^{-1}kpc^{-1}}$ at all the three radii mentioned. Although the choice of this  value is somewhat ad-hoc, nevertheless it helps us to predict the quantitative variation in the vertical density distribution for a higher gradient value.

The various other papers that study the variation of vertical velocity dispersion, as referred to in Section.1, do not separate the measured data into thin and thick disc contribution. Therefore, we do not use the gradient from  their data.

\subsubsection{Choice of other input parameters for the model}

The surface density values of the exponential stellar disc are calculated from the Galaxy mass model of Mera, Chabrier \& Schaeffer (1998), where the central surface density value is $\mathrm{\Sigma_{0}=640.9~M_{\odot}pc^{-2}}$ and the radial scale length is $R_{\mathrm{D}}$ =3.2 kpc.

For the multi-component system, we use the atomic hydrogen (HI) gas surface density values at $R$=6, 8.5 ,10 kpc as 4.6, 5.5 \& 5.5 $\mathrm{M_{\odot}pc^{-2}}$ respectively (Scoville \& Sanders 1987). We use $\sigma_{z}$ for HI gas as 8 $\mathrm{km~s^{-1}}$, and isothermal, at each of these radii, based on the values given by Spitzer (1978) for the Galaxy, and Lewis (1984) for nearly 200 face-on galaxies.

For the molecular hydrogen ($\mathrm{H_{2}}$) gas, the surface density values are measured to be 10.8, 2.1 \& 0.8 $\mathrm{M_{\odot}pc^{-2}}$ at $R$=6, 8.5, 10 kpc respectively (Scoville \& Sanders 1987) and $\sigma_{z}$ is taken to be 5 $\mathrm{km~s^{-1}}$ (Clemens 1985; Stark 1984), and isothermal, at each of these radii (see Table 1 (Sarkar \& Jog 2018) for all the surface density values).

For the dark matter halo we use the parameters as $R_{\mathrm{c}}=$5 kpc, and $V_{\mathrm{rot}} = \mathrm{220~km~s^{-1}}$ (see Section 2.1), as obtained by Mera , Chabrier \& Schaeffer (1998).

\section{Results}

The dynamical effect of non-isothermal dispersion of stars can be studied clearly in a system where the non-isothermal vertical pressure is balanced by the self-gravity of the stars-alone disc, which is the case considered first.

\subsection{Non-isothermal vertical structure for stars-alone disc}

\subsubsection{The vertical density distribution}

We apply the gradient in vertical stellar dispersion of +6.7 $\mathrm{km~s^{-1}kpc^{-1}}$ as observed (Hagen \& Helmi 2018) in Eq.(8), as discussed in the previous Section and solve Eq. (8) at $R$=6, 8.5, 10 kpc and show the numerical solutions $\rho(z)~vs.~z $ along with their corresponding isothermal solutions in Fig1. The isothermal solution is obtained by a sech$^2$ law (Eq.(5)) or by solving Eq.(8) on setting $C=0$, see Section 2.1.  
We also studied the non-isothermal effect using a higher dispersion gradient of +10 $\mathrm{km~s^{-1}kpc^{-1}}$ as discussed in sec 2.2.1 and show the resulting $\rho(z)$ distributions in Fig.1.

At each radius, the non-isothermal density distribution $\rho(z)$ is found to have a lower mid-plane density ($\rho_{\mathrm{0}}$) value and a higher scale height value, measured by the half width at half maximum  (HWHM) of the distribution, than the corresponding isothermal $\rho(z)$ distribution and therefore is more extended along the vertical direction. This effect is due to the higher vertical pressure in the non-isothermal case.
For example, at the solar radius, the $\rho_{\mathrm{0}}$ value is lower by 32.6 \% and the HWHM is higher by 37.1 \% than the isothermal case for the gradient of +6.7 $\mathrm{km~s^{-1}kpc^{-1}}$, therefore making the non-isothermal distribution  flatter along $z$. The reason is as follows.

\begin{figure*} 
\centering
\includegraphics[height=2.3in,width=3.2in]{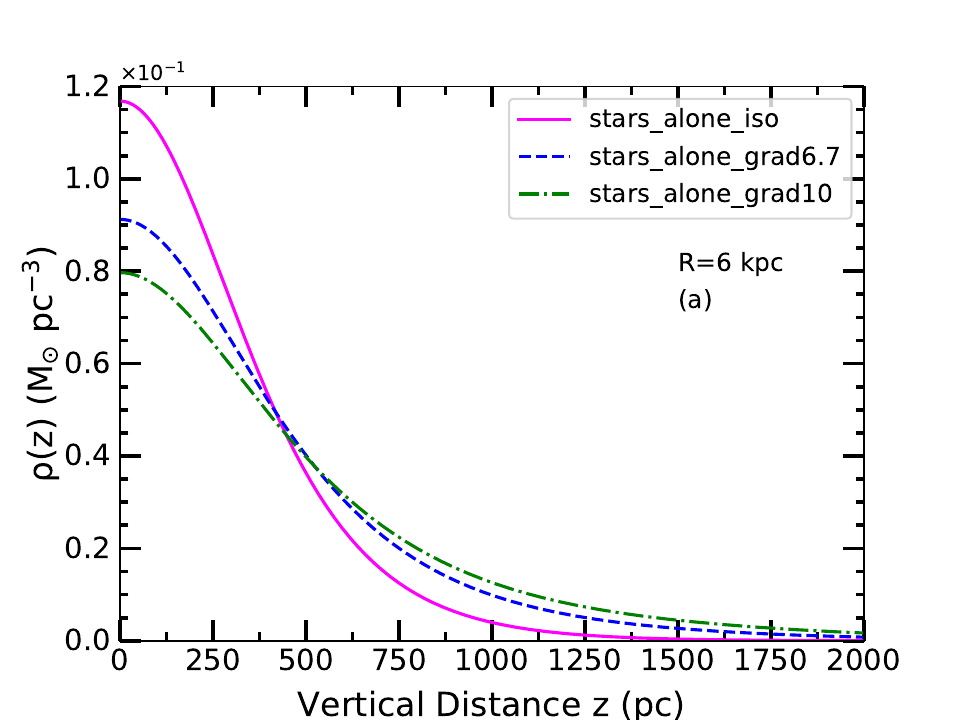}
\medskip
\includegraphics[height=2.3in,width=3.2in]{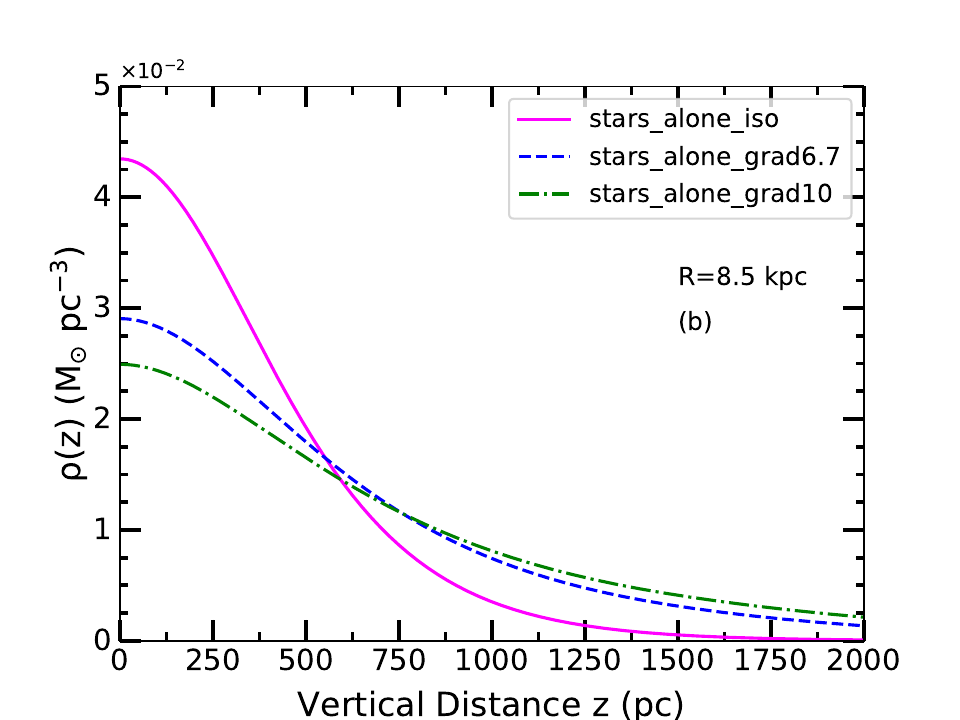}
\medskip
\includegraphics[height=2.3in,width=3.2in]{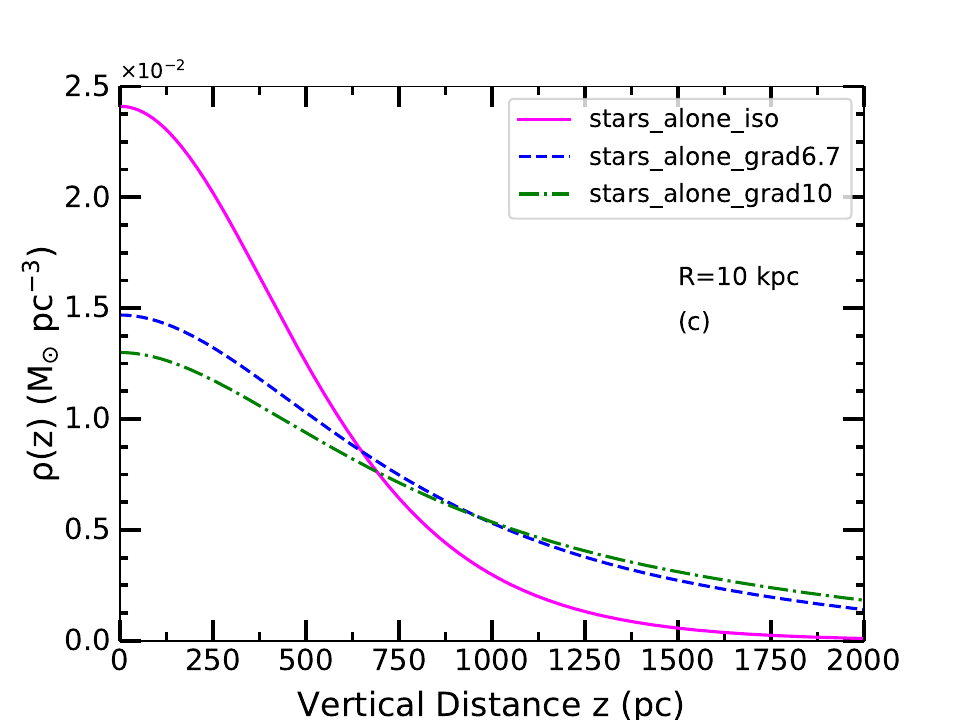}
\bigskip
\caption{Plot of resulting stellar vertical density distribution, $\rho(z)$ vs. $z$ at $R$= 6, 8.5 \& 10 kpc (a,b,c) respectively for a stars-alone disc. In each plot, the solid curve represents isothermal distribution. The dashed and the dashed-dotted curves represent non-isothermal distributions obtained using the dispersion gradient of +6.7 $\mathrm{km~s^{-1}kpc^{-1}}$ (derived from Hagen \& Helmi 2018) and a higher one, +10 $\mathrm{km~s^{-1}kpc^{-1}}$, respectively. The higher vertical pressure in the non-isothermal case results in a lower mid-plane density $\rho_{\mathrm{0}}$ and a higher scale height compared to the isothermal case, and thus gives rise to a more extended vertical distribution. This effect is more prominent with increasing gradient due to higher pressure and also with increasing radii due to the lower disc self-gravity.}
\label{label1}
\end{figure*}

In the non-isothermal case the dispersion increases with $z$ from its mid-plane value, at any radius, this makes the non-isothermal vertical pressure higher than the isothermal vertical pressure, at each $z$ height. Therefore the hydrostatic balance between the self-gravity and the non-isothermal vertical pressure makes the $\rho(z)$ distribution vertically more extended than the isothermal one. This effect will be stronger for a higher dispersion gradient. For example, due to the dispersion gradient of +10 $\mathrm{km~s^{-1}kpc^{-1}}$, the mid-plane density value at the solar radius decreases by 41.9\% and the HWHM value increases by 53.7\% than the isothermal case.

\begin{table} 
\begin{threeparttable}
\caption{Results for stellar mid-plane density values for a stars-alone disc} 
\label{table:1}
\centering
  \begin{tabular}{l l l l }      
\hline \hline
Radius & $\mathrm{\rho_{0,iso}}$ & $\mathrm{\rho_{0,non-iso}}$\tnote{a} & change  \\  
(kpc) & $\mathrm{(M_{\odot}\mathrm{pc^{-3}})}$ & $\mathrm{(M_{\odot}\mathrm{pc^{-3}})}$ & $(\%)$   \\  
\hline													 				
6.0	&	0.117	&	0.091	&	-22.2	\\ 
8.5	&	0.043	& 	0.029	&	-32.6	 \\
10.0	&	0.024	& 	0.015	&	-37.5	\\
\hline
\end{tabular}
\begin{tablenotes}\footnotesize
\item[a]The linear gradient with $z$ in $\sigma_{z}$ applied is +6.7 $\mathrm{km~s^{-1}~kpc^{-1}}$ (derived from Hagen \& Helmi 2018).
\end{tablenotes}
\end{threeparttable}
\end{table}

\begin{table} 
\begin{threeparttable}
\caption{Results for scale height (HWHM) values for a stars-alone disc}
\label{table:2}
\centering
\begin{tabular}{l l l l }      
\hline \hline
Radius & $\mathrm{HWHM_{iso}}$ & $\mathrm{HWHM_{non-iso}}$\tnote{a} & change  \\  
(kpc) & (pc) & (pc) & $(\%)$ \\  
\hline													 				
6.0	&	370.8	&	451.1	&	+21.6	\\ 
8.5	&	456.7	&	626.2	&	+37.1   \\
10.0	&	515.4	&	763.1	&	+48.1	\\
\hline
\end{tabular}
\begin{tablenotes}\footnotesize
\item[a]The linear gradient with $z$ in $\sigma_{z}$ applied is +6.7 $\mathrm{km~s^{-1}~kpc^{-1}}$ (derived from Hagen \& Helmi 2018).
\end{tablenotes}
\end{threeparttable}
\end{table}

We also note from Fig. 1, that the non-isothermal effect becomes more prominent with increasing radii for a given dispersion gradient. This happens as the self-gravity of the stellar disc decreases with increasing radii and hence the non-isothermal vertical pressure can affect the distribution more.

\begin{figure*}
\centering
\includegraphics[height=2.3in,width=3.2in]{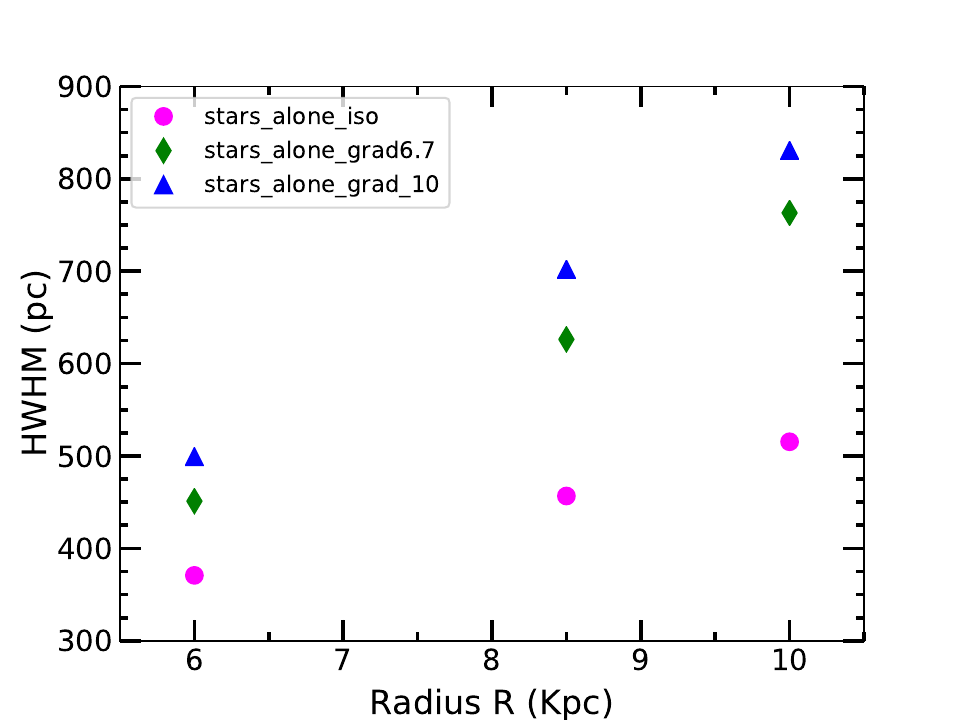}  
\caption{Plot of HWHM values of the resulting non-isothermal $\rho(z)$ distributions as a function of radius for the stars-alone disc along with the corresponding isothermal values. The gradients applied in $\sigma_{z}$ are +6.7 $\mathrm{km~s^{-1}kpc^{-1}}$ \& +10 $\mathrm{km~s^{-1}kpc^{-1}}$ respectively. The non-isothermal dispersion gives an extended distribution with a higher scale height that flares (increases) with radius, and this flaring is higher for higher dispersion gradient value.}
\label{label2}
\end{figure*}

In the outer disc, well beyond the solar neighbourhood region, the self-gravity of the stellar disc is quite low. Therefore the non-isothermal effect will make the vertical distribution much more extended. Such an extended vertical disc, now can exert less self-gravitational force at the high $z$ or the extended part of the distribution, compared to the isothermal one and therefore will be significantly more susceptible to any external tidal perturbation. So it may be interesting for observers to study the vertical stellar dispersion values in the outer disc as a function of $z$, this would enable our model to be applied to this region to theoretically obtain the vertical stellar density profiles. 

We give the resulting mid-plane density ($\rho_{\mathrm{0}}$) values (upto 3rd decimal) and the scale height (HWHM) values of the non-isothermal distributions (obtained for $\mathrm{d}\sigma_{z}/\mathrm{d}z=\mathrm{+6.7~km~s^{-1}kpc^{-1}}$) in Table 1 \& 2 respectively and compare them with the corresponding isothermal results. This gives us a quantitative idea of how important it is to consider the non-isothermal velocity dispersion in the determination of the vertical structure for a stellar disc at different Galactic radii, and hence it needs to be included in the dynamical modelling of the Galaxy.

Fig. 2 shows the variation of HWHM at $R$=6,8.5,10 kpc, due to all the applicable cases (isothermal, and non-isothermal with the different gradients), discussed so far. Fig. 2 shows a flaring trend with radius, with the flaring being higher in the non-isothermal cases with progressively higher values for the higher dispersion gradient considered.
 
We note an important point regarding the procedure used to solve Eq.(8) as given in Section 2.1. We integrate the solution $\rho(z)$ along $z$ to obtain the surface density value and compare it with the observed value to satisfy the boundary condition. But as the $z$ range for integration is increased for the given surface density, the solution $\rho_{0}$ also changes. Therefore we limit the $z$ range for solving the equation, at a value such that the mid-plane density does not change by more than 2$\%$ if $z$ range is increased further by a small range $\mathrm{\sim 1~kpc}$. We denote this limit as $z_{max}$ and note that its value depends on radius and dispersion gradient. We note that the gradient value of +6.7 $\mathrm{km~s^{-1}kpc^{-1}}$, used here is obtained from the observed dispersion data given up to $z$=1.5 kpc only (Hagen \& Helmi 2018), but $z_{max}$ is found to be higher than this at all radii. Since we do not know the behavior of $\sigma_{z}$ for $z$ greater than 1.5 kpc, therefore we assume the dispersion to get saturated and then remain constant at its highest value reached at $z$=1.5 kpc, at all $z$ values beyond this. We apply the same technique, as discussed above, while solving with the gradient of +10 $\mathrm{km~s^{-1}kpc^{-1}}$ too.

We also tried out the observed gradient of +7.2 $\mathrm{km~s^{-1}kpc^{-1}}$ (Jing et al. 2016), as discussed in the sec 2.2.1. Since the two observed gradient values are very close to each other, the results are found to differ only very slightly from each other (e.g. to within 3\% at 8.5 kpc) and we do not show them in the plots.

\subsubsection{Fitting single and double $\sech^{2}$ profiles to the density distribution}

It has long been a tradition to describe the observed vertical mass or luminosity density distribution of stars using a $\sech^{2}$ profile following Spitzer's law (Eq. 5, sec.2.1). In recent papers in the literature, the stellar vertical density distribution of the Milky Way is often described by a double $\sech^{2}$ profile as applied to observed data (Bovy 2017; Ferguson, Gardner \& Yanny 2017; Wang et al. 2018), as well as to results from simulations (Ma et al. 2017), mainly to account for the thin and the thick stellar discs. The scale heights of the two discs are also determined using the best-fit parameters of the thin plus thick disc fit. This is routinely done based on the observed fact that the Galaxy has a thick disc of stars (Gilmore \& Reid 1983) which is photometrically, kinematically and chemically well-defined and therefore this thick disc is a physically distinct entity.

The same method of a double disc fitting is routinely followed for the external edge-on galaxies too (Matthews 2000; Yoachim \& Dalcanton 2006 ; Comer$\mathrm{\acute{o}}$n et al. 2011). The vertical profile in these galaxies is often found to deviate from a single $\sech^{2}$ function, specially at high $z$ and the excess flux at high $z$ makes the observed distribution look like a "wing" w.r.t the fitted single $\sech^{2}$ function. To account for this wing or high $z$ part of the distribution, another $\sech^{2}$ profile representing a thicker disc is added in the fitting in the above papers.

Now, we try to check if the non-isothermal $\rho(z)$ distribution of a single stellar disc, instead of two genuine isothermal discs, can explain the above behavior. Therefore, we fit the $\rho(z)$ distributions, calculated using the vertical dispersion gradient of +6.7 $\mathrm{km~s^{-1}kpc^{-1}}$ \& +10 $\mathrm{km~s^{-1}kpc^{-1}}$, at the solar radius by single and double $\sech^{2}$ profiles using the following functions respectively
\begin{equation}
\rho(z)=\rho_{0}\sech^{2}(z/z_{0})\label{eq:17} 
\end{equation}
\noindent
\begin{equation}
\rho(z)=\rho_{0,\mathrm{thin}}\sech^{2}(z/z_{\mathrm{thin}})+\rho_{0,\mathrm{thick}}\sech^{2}(z/z_{\mathrm{thick}})\label{eq:18}
\end{equation}

\noindent where we have taken ($\rho_{\mathrm{0}}$, $z_{0}$) as best-fit parameters for the single $\sech^{2}$ profile and ($\rho_{0,\mathrm{thin}}, z_{\mathrm{thin}}, \rho_{0,\mathrm{thick}}, z_{\mathrm{thick}}$) as best-fit parameters for a double (thin plus thick) disc profile respectively. The best-fit functions along with the respective density distributions are shown in Fig.3. 

\begin{figure*}
\centering
\includegraphics[height=2.3in,width=3.2in]{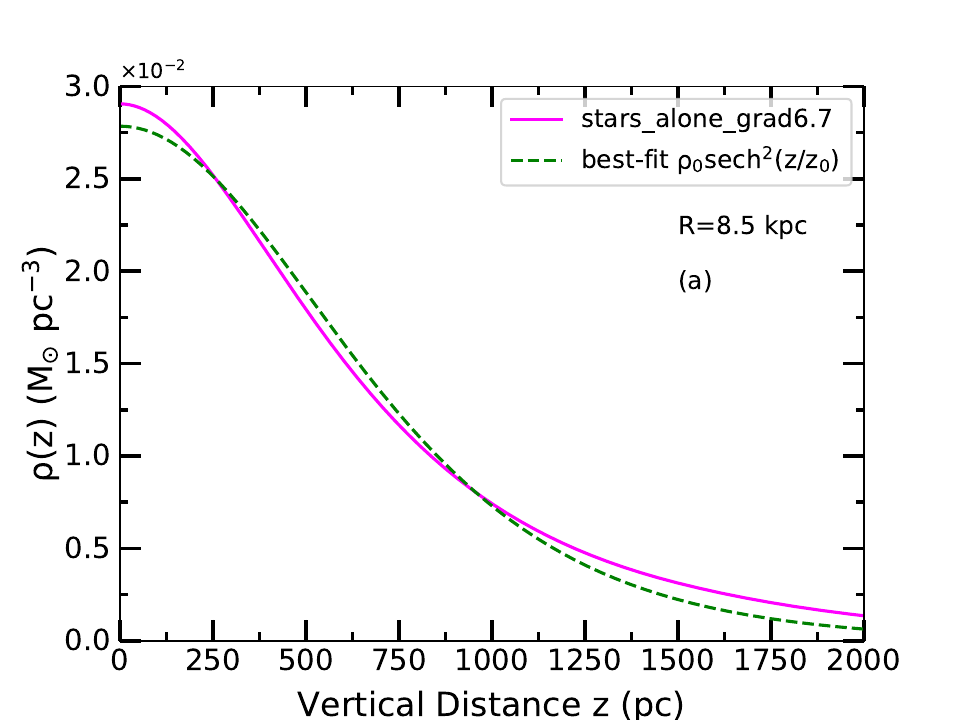}   
\medskip
\includegraphics[height=2.3in,width=3.2in]{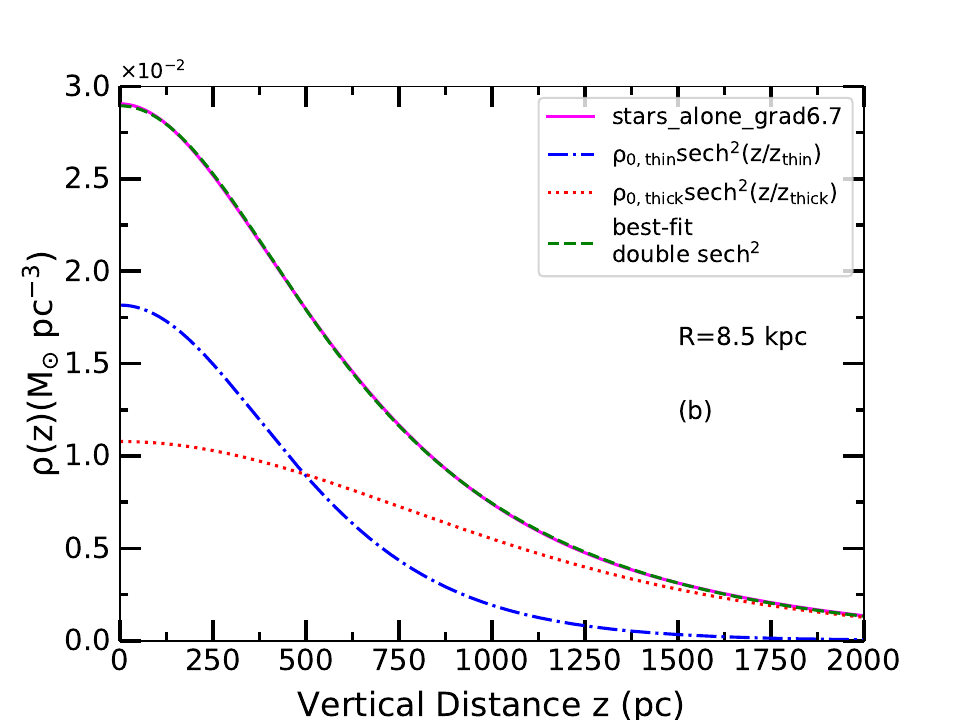}  
\bigskip
\includegraphics[height=2.3in,width=3.2in]{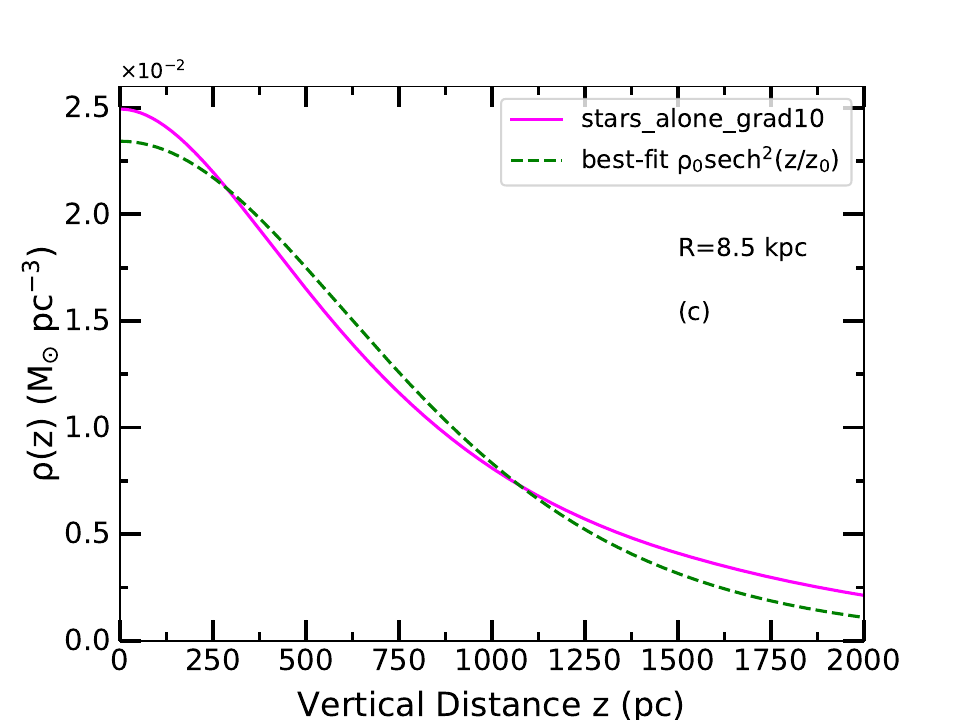}  
\medskip
\includegraphics[height=2.3in,width=3.2in]{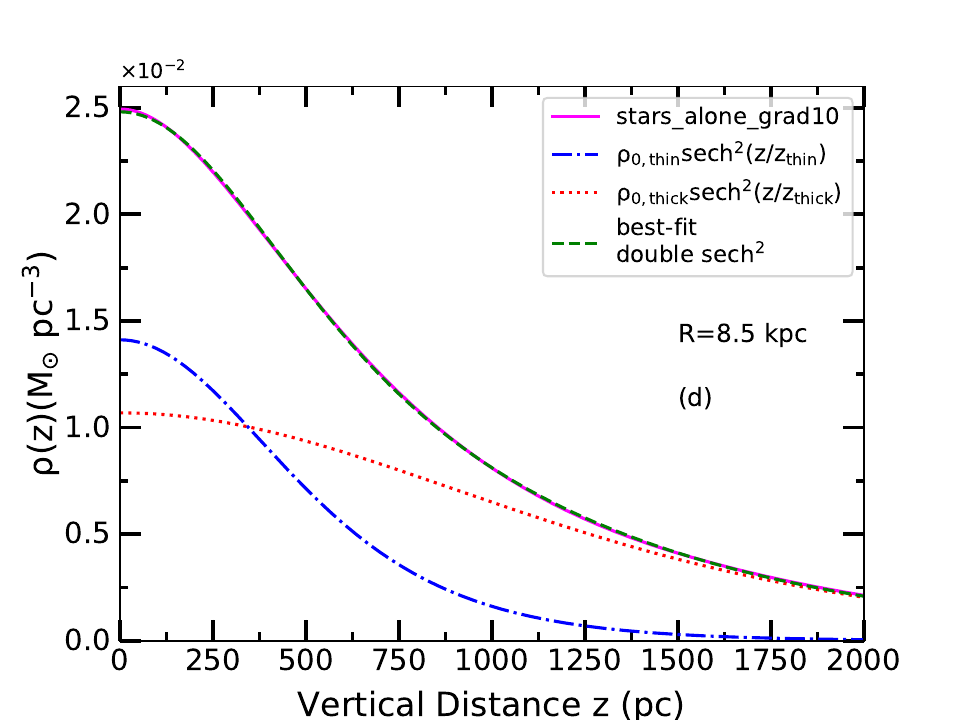}  
\bigskip
\caption{Fitting of single and double $\sech^{2}$ profiles to the non-isothermal $\rho(z)$ distribution of stars for the stars-alone case at $R$=8.5 kpc. The left panel (a,c) shows the best-fitting single $\sech^{2}$ profiles to the $\rho(z)$ distributions obtained using the dispersion gradients of +6.7 \& +10 $\mathrm{km~s^{-1}kpc^{-1}}$ respectively. The function gives a poor fit to the density distribution at all $z$, especially at the "wing" or high $z$ part of the distribution. This behavior is more prominent for the higher gradient. The right panel (b,d) shows the best-fit double $\sech^{2}$ profiles consisting of a "thin plus thick disc", which give a good fit to the density distribution at all $z$. The density distribution in the "thin \& thick discs" of the double disc profile are also plotted. This shows how an observer may mis-interpret a non-isothermal single stellar disc distribution as a superposition of two separate $\sech^{2}$ or isothermal discs.}
\label{label3}
\end{figure*}

It is clearly seen that the best-fitting single $\sech^{2}$ function gives a poor fit to the $\rho(z)$ distribution, especially at high $z$ or the wing region. The deviation is even higher with the dispersion gradient of +10 $\mathrm{km~s^{-1}kpc^{-1}}$. On the other hand, the double $\sech^{2}$ disc profile fits our results very well at all $z$ for both the gradients considered. This can lead to the mis-interpretation that the vertical distribution is determined by a superposition of two separate discs, a thin disc and a thick disc. This is more likely to happen for the external edge-on galaxies, where an independent measure of thin and thick disc components may not be available. Therefore, we caution that if an observed vertical density or luminosity profile shows wing at higher $z$ and a single $\sech^{2}$ function is found to give a poor fit to the data, it may well be due to a \textit{non-isothermal $\rho(z)$ distribution of stars of a single stellar disc}. Thus evoking a second thicker disc to fit the observed profile could be redundant. 

Similarly, for the Milky Way, a part of the deviation of the vertical profile from a single $\sech^{2}$ function, especially at high $z$, can arise due to the observed non-isothermal dispersion as well, apart from the genuine thick disc contribution. This would affect the calculation of the thin and thick disc parameters obtained by doing a double-disc fit to the data. We suggest that future work on this topic should take account of this point.

\subsection{Results for the realistic multi-component system}

\subsubsection{The vertical density distribution}

\begin{figure*}   
\centering
\includegraphics[height=2.3in,width=3.2in]{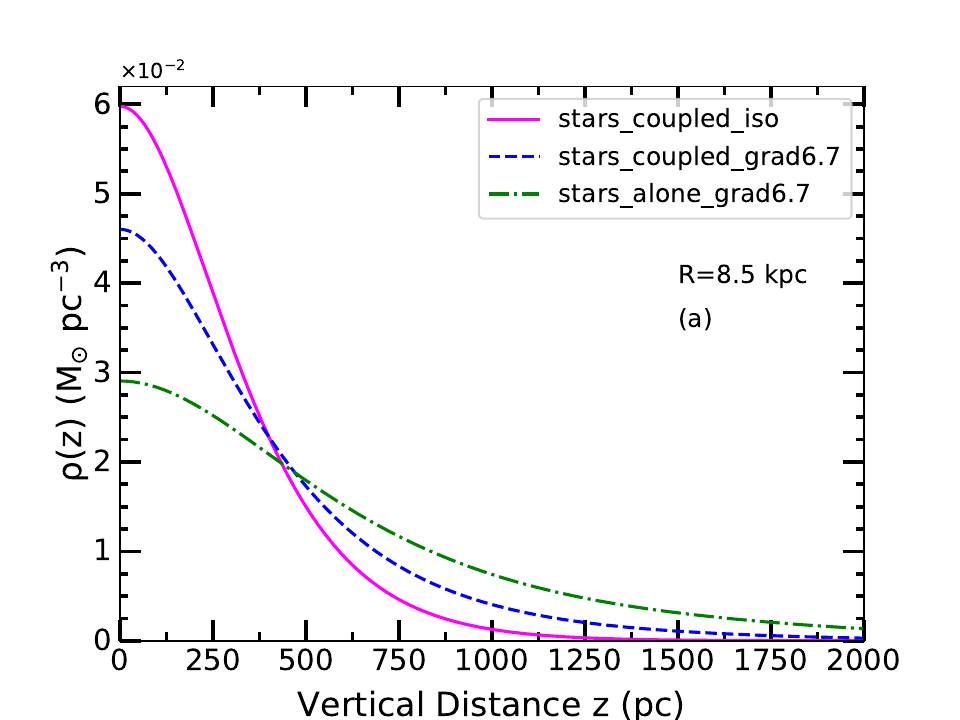}    
\medskip
\includegraphics[height=2.3in,width=3.2in]{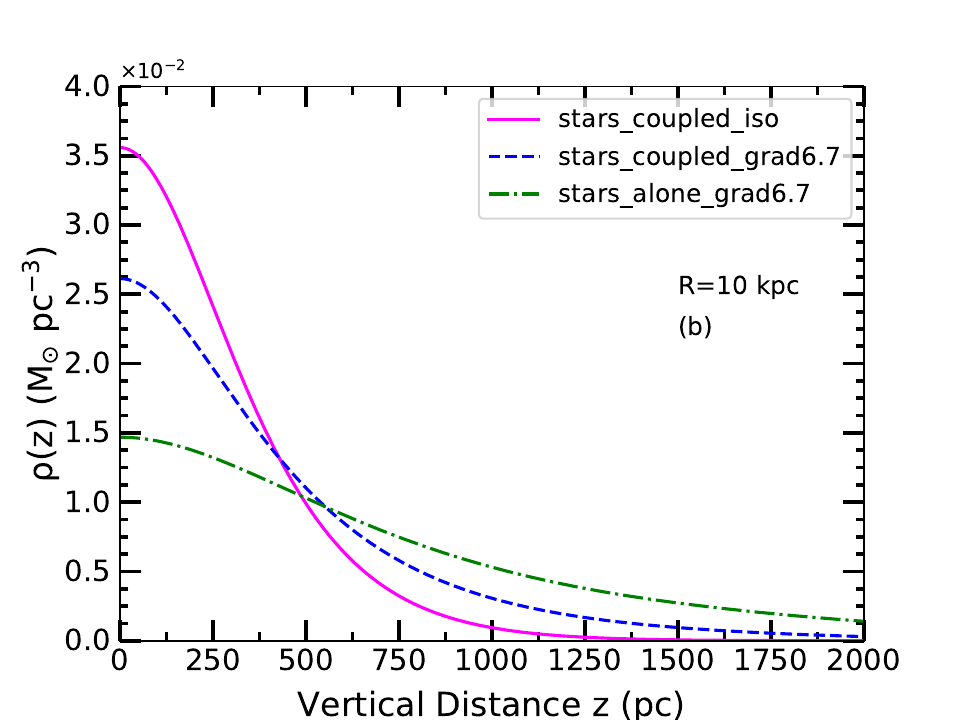}   
\caption{The resulting non-isothermal vertical density distribution of stars is shown at $R$=8.5 \& 10 kpc (a,b respectively) for the stars-alone case (dashed-dotted curves) and for the stars plus gas plus halo coupled system (dashed curves), obtained using the dispersion gradient of $\mathrm{d}\sigma_{z}/\mathrm{d}z=+6.7\mathrm{~km~s^{-1}kpc^{-1}}$ . The non-isothermal distribution in the coupled case is more constrained toward the mid-plane by the gravity of gas and the dark matter halo compared to the stars-alone case. But it still remains more extended than the isothermal stellar distribution of the coupled system (the solid curves.)}
\label{label4}
\end{figure*}

The multi-component system of gravitationally coupled Galactic disc of stars and gas in the field of dark matter halo was treated for the isothermal case by Sarkar \& Jog (2018) where the vertical distribution of stars was shown to be constrained mainly by the gas gravity in the inner disc and the dark matter halo gravity in the the outer disc, respectively. This effect is opposite to the extended distribution caused by non-isothermal dispersion as studied above (Section 3.1). We next consider the non-isothermal, multi-component case and compare the results to the non-isothermal stars-alone case, as well as to an isothermal, multi-component coupled case.

To treat the non-isothermal coupled case, we solve the Eq.(12,13) together using dispersion gradient of +6.7 $\mathrm{km~s^{-1}kpc^{-1}}$ at $R$=8.5 \& 10 kpc, following the methods discussed in sec 2.1 and show the resulting distributions in Fig.4. The non-isothermal $\rho(z)$ distribution for the coupled system is now found to be more constrained toward the mid-plane than the corresponding non-isothermal stars-alone distribution. For example, the mid-plane density values at $R$=8.5, 10 kpc are now 0.046 \& 0.026 $\mathrm{M_{\odot}pc^{-3}}$ respectively, which are higher by 58.6\% and 73.3\% from the corresponding stars-alone cases (see Table 1). The HWHM values are respectively 396.9 \& 431.7 pc, that are lower by 36.6\% \& 43.4\% compared to the stars-alone cases (see Table 2). Similarly, the non-isothermal effect due to the dispersion gradient of +10 $\mathrm{km~s^{-1}kpc^{-1}}$ is also less here than the corresponding stars-alone case.

We also plot the density distribution for the isothermal, multi-component case in Fig.4 (obtained by solving Eq.12 \& 13 with $C=0$; also see Sarkar \& Jog (2018)). Interestingly, although, gas and halo have constrained the non-isothermal distribution here compared to the stars-alone case, nevertheless it remains more extended than the isothermal solutions of the coupled system. For example, $\rho_{0}$ at R=8.5 kpc, for the gradient of +6.7 $\mathrm{~km~s^{-1}kpc^{-1}}$, is less than the $\rho_{0}$ of the isothermal coupled case by 23\%. Thus, the effect of non-isothermal dispersion is opposite to the constraining effect of the gas and halo gravity and dominates over it.

We note that the corresponding gas (HI and $\mathrm{H_{2}}$) distribution is affected by less than 8$\%$ at $R$=8.5 kpc taking a non-isothermal stellar velocity dispersion. This is because in the coupled case, the gas distribution is nearly independent of the stellar velocity dispersion (and vice versa), this is similar to the result shown for the isothermal case by Narayan \& Jog (2002).

To solve for the coupled case, we used the same $z_{max}$ for integration as used for the corresponding stars-alone cases (Section 3.1.1). Since the coupled cases are less extended than the stars-alone cases, this ensures that there is at least the same level of accuracy or even higher in the coupled system results at each radius, than in the stars-alone cases.

\subsubsection{Fitting single and double $\sech^{2}$ profiles to the density distribution}

\begin{figure*}
\centering
\includegraphics[height=2.3in,width=3.2in]{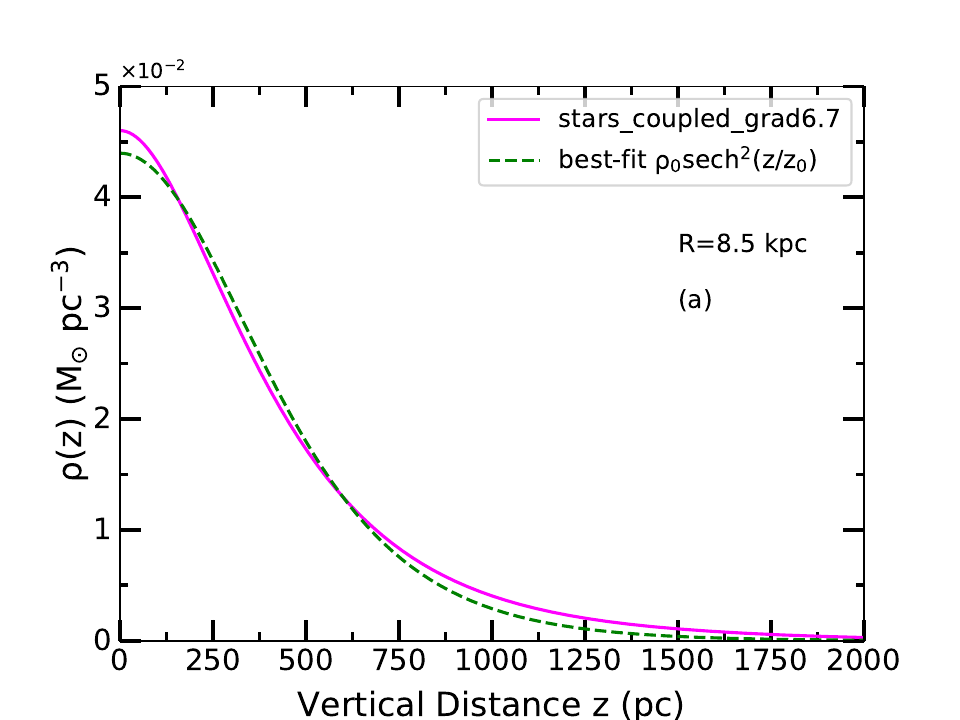}  
\medskip
\includegraphics[height=2.3in,width=3.2in]{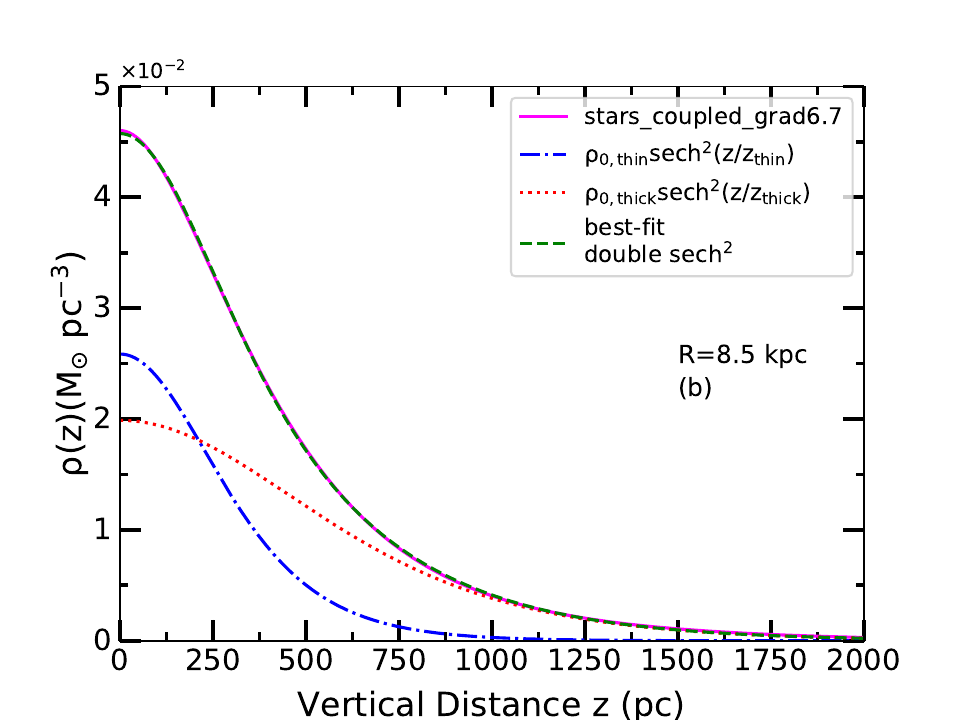}  
\caption{Fitting of single and double $\sech^{2}$ profiles to the non-isothermal $\rho(z)$ distribution of stars for the coupled stars+gas+halo system at $R$=8.5 kpc obtained for the dispersion gradient of +6.7 $\mathrm{km~s^{-1}kpc^{-1}}$. The best-fitting single $\sech^{2}$ function fails to reproduce the $\rho(z)$ distribution at all $z$ (panel a), more at the "wing" or high $z$ part of the distribution. The double $\sech^{2}$ profile consisting of a "thin+thick disc", gives a good fit to $\rho(z)$ (panel b) at all $z$. The density distribution in the "thin \& thick discs" of the double disc profile are also plotted. This shows how an observer may mis-interpret a non-isothermal single stellar disc distribution in the coupled case as arising due to a superposition of two separate $\sech^{2}$ or isothermal discs.}
\label{label5}
\end{figure*}

We choose the solution for the density distribution for the multi-component system at the solar radius ($R$=8.5 kpc), obtained using the gradient value of $\mathrm{d}\sigma_{z}/\mathrm{d}z=+6.7~\mathrm{km~s^{-1}kpc^{-1}}$ and fit a single and a double $\sech^{2}$ profiles (Fig. 5), as was done earlier in Sec 3.1.2 for the stars-alone case. As usual, the thin+thick disc profile reproduces the distribution successfully at all $z$, unlike the single $\sech^{2}$ profile. Thus our results show that, even for the realistic system, evoking a thicker disc to get a good fit to the observed data, specially to the "wing" part, may be redundant. 

We note that this deviation in the non-isothermal coupled disc arises out of two contrasting effects- the constraining effect of gas and halo on the distribution of stars and the opposite trend of extension of the stellar vertical distribution caused by the non-isothermal dispersion of stars. These two effects dominate at different $z$ heights. Hence the deviation at different $z$ will be a combination of these two effects. In general, the effect of non-isothermal dispersion is seen to be the dominant factor that determines the net density distribution, hence its deviation from $\sech^{2}$ profile.

Again, we note that we do not rule out the contribution of the genuine thick disc in causing this deviation, for the Milky Way. However, this finding of a redundant second disc would be more applicable for the external edge-on galaxies where a thin plus thick disc fitting is routinely carried out in order to get a good fit to the observed vertical profiles.  

We note that a similar mis-representation of a flaring disc attributed to a second, thick disc may also occur in an isothermal case, as was shown for the low surface brightness galaxy UGC 7321 (Sarkar \& Jog 2019) where the observed increase in disc thickness with radius was shown to be explained by a self-consistent, multi-component disc plus halo model. Thus Sarkar \& Jog (2019) had argued that a thick disc often invoked in the literature for such galaxies is not necessary.

\subsection{Fitting a $\sech^{2/n}$ function to the density distribution}
It has been realized that a simple $\sech^{2}$ does not give a good fit to the observed density profiles of galactic discs. In order to get a better fit to the data, van der Kruit (1988) suggested an analytical form, given below, to describe the observed vertical density distribution of stars

\begin{equation}
\rho(z)=\rho_{e}2^{-2/n}\sech^{2/n}(nz/2z_{e})\label{eq:19}
\end{equation}

\noindent where the exponent value $2/n$ is considered to be an indicator of the shape of the profile and $z_{e}$ is a measure of scale height. Here $n$=1, 2 \& $n$$\rightarrow \infty$ denote the $\sech^{2}$, $\sech$, \& exponential vertical profiles respectively. We also note that $\rho_{e}2^{-2/n}$ denotes the mid-plane density. We caution that this function has no clear physical basis but is used because it is a handy, analytical criterion to compare the results for different galaxies.

We apply this function to fit the non-isothermal stellar distribution of the coupled system at $R$=8.5 kpc and 10 kpc, obtained using the gradient of +6.7 $\mathrm{km~s^{-1}kpc^{-1}}$. The best-fit function is found to reproduce the model data fitting range well at all $z$.

However, the best-fit value of the exponent $2/n$ changes with the fitting range ($\Delta z$) used, i.e, the value first increases, then decreases and finally saturates. Hence, the best-fit value of the exponent is not robust. We show the best-fit values of $2/n$ as a function of $\Delta z$ in Fig. 6. We also note that the values always remain less than 1 at both the radii.

\begin{figure*}
\centering
\includegraphics[height=2.3in,width=3.2in]{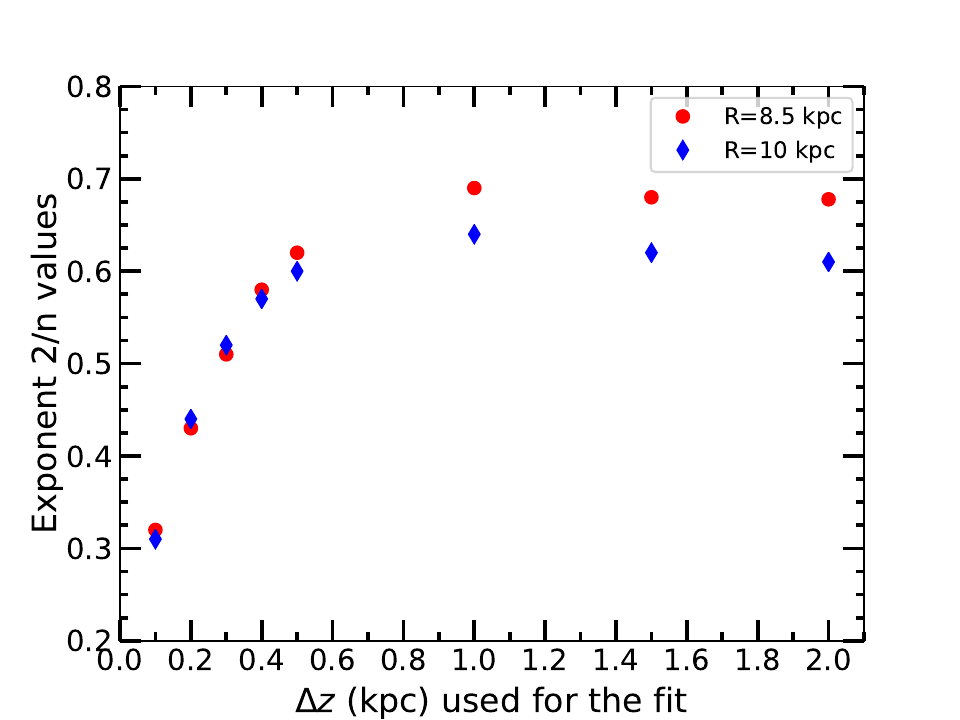}   
\caption{Plot of best-fit value of the exponent $2/n~vs.\Delta z$, when a $\sech^{2/n}$ type function (Eq.19) is fitted to the non-isothermal vertical density distribution of stars obtained for the stars plus gas plus dark matter halo system at $R$=8.5 \& 10 kpc for the dispersion gradient of +6.7 $\mathrm{km~s^{-1}kpc^{-1}}$. The $2/n$ values change with the fitting range $\Delta z$, but remain less than 1.}
\label{label6}
\end{figure*}

Interestingly, in our earlier work in Sarkar \& Jog (2018), we had found a different behavior in $2/n~vs.~z$ when the function was fitted to the $\textit{isothermal}$ distribution of stars of the coupled system. The $2/n$ values were found to increase with $\Delta z$ and reach a value greater than 1 and even 2, specially in the outer disc (for a detailed discussion and explanation see Sarkar \& Jog, 2018). Hence, we conclude that it is the non-isothermal nature of the distribution that causes the trend, as shown in Fig.6. We further checked that this is also evident for the non-isothermal stars-alone case (treated in Section 3.1). 
 
In summary, this result brings out the point that a single but non-$\sech^{2}$ profile can fit the non-isothermal $\rho(z)$ distribution over a fixed fitting range, and thus it may not be necessary to use a double $\sech^{2}$ distribution to fit this.

\section{Effect of non-isothermal dispersion on determination of Oort limit}

The Oort limit is one of the most important physical quantities studied in Galactic dynamics. It is the dynamical mass volume density, defined at the mid-plane at solar radius. It represents a measure of distribution of mass of the Galactic disc near the sun. It was measured to be 0.15 $\mathrm{M_{\odot}pc^{-3}}$ in the pioneering work by Oort (1960). It was derived from the measurement of vertical Galactic force field, calculated using the observed star counts and vertical velocity dispersion of tracer stars using the isothermal approach. 

Although Oort considered the stellar population in the solar neighbourhood to be isothermal, a superposition of more than one such isothermal population was required in the calculation to satisfy the observational constraints. Later papers have done a redetermination of the Oort limit value, using more precise observed data for star counts and kinematics (Bahcall 1984b,c; Kuijken \& Gilmore 1989; Bahcall et al. 1992; Cr$\mathrm{\acute{e}}$z$\mathrm{\acute{e}}$ et al. 1998; Holmberg \& Flynn 2000) under the isothermal approximation. These papers did not treat non-isothermal dispersion as a physical parameter and did not obtain a self-consistent vertical distribution for such a case, as we have done in this paper.

Now we use our theoretical model results to study the effect of non-isothermal dispersion on the determination of the Oort limit value or the total dynamical mid-plane density. The non-isothermal vertical distribution of stars for the standard observed dispersion gradient of +6.7 $\mathrm{km~s^{-1}kpc^{-1}}$ (Hagen \& Helmi 2018), obtained for the gravitationally coupled stars plus gas plus dark matter halo system at the solar radius, gives $\rho_{0}$ as 0.046 $\mathrm{M_{\odot}pc^{-3}}$. The distributions for HI, $\mathrm{H_{2}}$, that are also obtained simultaneously, give $\rho_{0}$ values as 0.017 \& 0.011 $\mathrm{M_{\odot}pc^{-3}}$ respectively. The dark matter halo contribution at the mid-plane at R=8.5 kpc is obtained using Eq.(14) to be 0.009 $\mathrm{M_{\odot}pc^{-3}}$ . Thus the total $\rho_{\mathrm{0,noniso}}$ for stars+gas+halo system, obtained from our self-consistent theoretical model solutions turns out to be 0.083 $\mathrm{M_{\odot}pc^{-3}}$. The isothermal solutions obtained by Sarkar \& Jog (2018), give the total mid-plane density as 0.099 $\mathrm{M_{\odot}pc^{-3}}$.

Thus our results show that the total mid-plane density determined by considering the non-isothermal dispersion is less than the one calculated for the isothermal case by 16 $\%$ which is significant. We note that although the total $\rho_{0,\mathrm{noniso}}$ obtained, falls within the range of the measured values ($\sim\mathrm{(0.076-0.2) M_{\odot}pc^{-3}}$) in literature, still we do not treat this as a redetermination of the "true" Oort limit value. This is because, its accurate measurement should include the coupling between vertical and radial motions (for example, see Sarkar \& Jog 2020), which is beyond the scope of this paper. Instead, our aim is to emphasize that the self-consistent treatment in the non-isothermal case can affect the measurement of the Oort limit value and therefore this feature should be included in its determination.

\section{Limit of validity of isothermal assumption}

\begin{figure*}
\centering
\includegraphics[height=2.3in,width=3.2in]{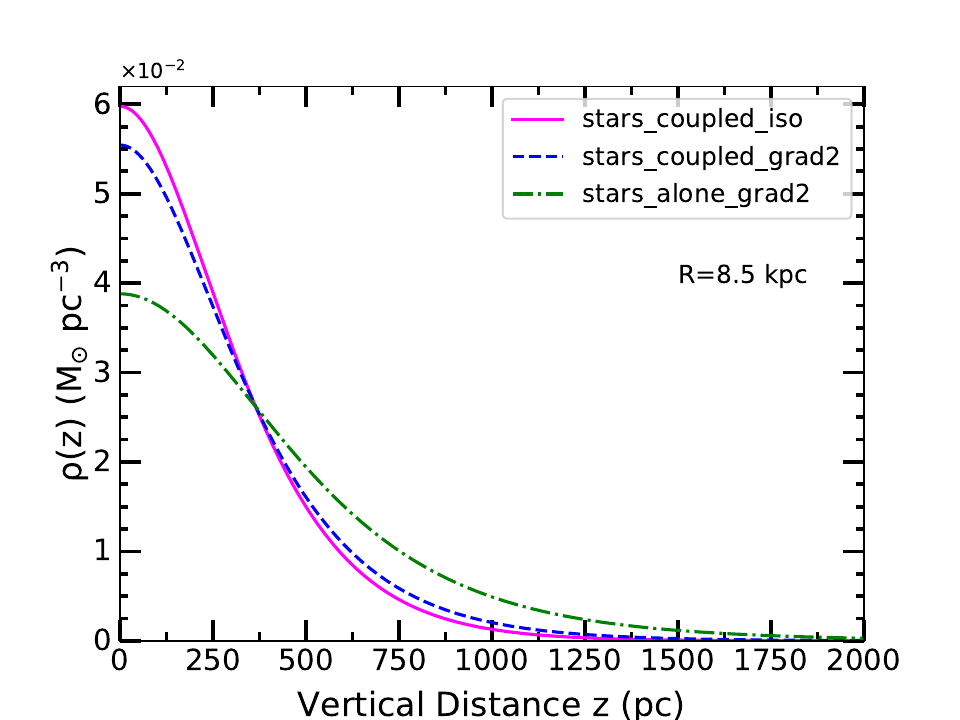}   
\caption{The resulting non-isothermal $\rho(z)$ distribution of stars at $R$=8.5 kpc for a small dispersion gradient of +2 $\mathrm{km~s^{-1}kpc^{-1}}$ for the stars-alone case (the dashed-dotted curve) and for the stars plus gas plus halo coupled system (the dashed curve) along with the isothermal coupled case (the solid curve). The non-isothermal distribution for the coupled system for this low dispersion gradient is now close to the distribution of the isothermal coupled case and therefore the isothermal assumption of the stellar disc may be reasonably valid for such low dispersion gradient.}
\label{figure7}
\end{figure*}

The vertical structure is routinely studied assuming the disc to be an isothermal disc, for simplicity (Section 1). We have shown that at the observed dispersion gradient of +6.7 $\mathrm{km~s^{-1}kpc^{-1}}$ the resulting vertical stellar disc distribution is affected substantially. The mid-plane density values change by -32.6 \% for a stars-alone disc and by -23.0 \% in the coupled system at the solar radius. It would be interesting to see upto which gradient value can we still consider the disc to be isothermal. We find that the values of the gradient equal to +2, +3, +4 \& +5 $\mathrm{km~s^{-1}kpc^{-1}}$ cause a change of -9.3\%, -14.0\%, -20.9\% \& -25.6\% in the mid-plane density values respectively for the stars-alone case. These changes are lower in a multi-component disc and are equal to -8.3\%, -11.7\%, -15.0\% \& -18.3\% respectively from the corresponding isothermal coupled system. Thus for the Galaxy, and presumably for other typical large spiral galaxies, a disk with a gradient value up to +2 $\mathrm{km~s^{-1}kpc^{-1}}$ could be effectively considered to be isothermal in the realistic multi-component case, for simplicity of the calculations. We show the resulting distributions with this gradient at the solar radius in Fig. 7. 
At the same time this work shows that even a small gradient $> 2~\mathrm{km~s^{-1}kpc^{-1}}$ can start to show its effect and the resulting density distribution shows a discernible deviation from the $\sech^{2}$ distribution especially for the stars-alone case. Thus we caution that the $\sech^{2}$ law that is traditionally used for its convenient, analytical form may not be rigorously correct. 

\medskip

\section{Conclusions}

We have studied the self-consistent vertical density distribution of stars in the Galactic thin disc when the vertical velocity dispersion is taken to be non-isothermal. We applied the typical observed dispersion gradient of +6.7 $\mathrm{km~s^{-1}kpc^{-1}}$ to the vertical velocity dispersion, as well as a higher value (+10 $\mathrm{km~s^{-1}kpc^{-1}}$). We illustrate the effects clearly, using a stars-alone disc first and then for completeness, study it for a multi-component gravitationally coupled disc of stars plus gas in the field of dark matter halo. This treatment yields a rich set of results, which are summarized below:

\noindent 1. The non-isothermal density distribution at any radius has less mid-plane density, and a higher HWHM value than the corresponding isothermal distribution, due to higher vertical pressure. Therefore it appears as a more vertically extended distribution, with changes in both these quantities being $\sim$ 35\% at the solar radius for a stars-alone disc. These effects become more important with increasing radii and increasing gradient values.

\noindent 2. In a realistic multi-component system the results show a similar trend, except here the inclusion of gas and dark matter halo gravity tends to constrain the distribution toward the mid-plane while the non-isothermal dispersion has an opposite effect. Therefore the non-isothermal coupled case is less extended than the stars-alone non-isothermal case, but still remains extended compared to the multi-component isothermal case. Thus a realistic, multi-component system is more robust and is able to withstand a high velocity dispersion gradient and is thus less likely to be disturbed by an external tidal encounter.

\noindent 3. We show that a non-isothermal distribution shows deviation from a single $\sech^{2}$ function, more at extended high $z$ part or the "wing" and is fitted well by a double $\sech^{2}$ function. Normally a second, thick disc is invoked in the literature to explain this, for the external edge-on galaxies, which thus is not necessary. Even the case when there is a genuine second disc present, as in the Milky Way, the net distribution and hence the disc parameters as obtained by the thin plus thick disc fitting to the data are likely to be affected by the non-isothermal dispersion. Hence the non-isothermal effect should be included in analysing the observed vertical profiles.

\noindent 4. We find that the non-isothermal dispersion lowers our theoretical estimate of the total mid-plane density (that is, the Oort limit) for the multi-component disc plus halo model, by $16 \%$ compared to the isothermal case. We note that we do not aim to provide a re-determination of the Oort limit value, instead we emphasize the importance of including non-isothermal velocity dispersion of stars in the calculation of the Oort limit.

\noindent 5. We check that for a stars-alone disc, even a smaller gradient of 2-3 $\mathrm{km~s^{-1}kpc^{-1}}$ can affect the mid-plane density values by ($\sim 9-14 \%$) compared to the isothermal case. Thus we caution that the $\sech^{2}$ law used traditionally for its convenience may not be rigorously correct.

Thus, we have shown that the non-isothermal stellar vertical velocity dispersion has important dynamical effect on the vertical disc structure and hence should be included in the future dynamical modelling of a galactic disc.

\noindent \textbf{Acknowledegments}:
We thank the anonymous referee for constructive comments, and for asking us the question about the possible effect of non-isothermal velocity dispersion along the vertical direction on the asymmetric drift in the disc. S.S. thanks CSIR for a fellowship, and C.J. thanks the DST for support via J.C. Bose fellowship (SB/S2/JCB-31/2014).

\medskip

\noindent \textbf{Data availability}:
The data underlying this article will be shared on reasonable request to the corresponding author.

\bigskip

\noindent {\bf {References}}
\medskip

\noindent Aumer M., Binney J., Sch$\ddot{\mathrm{o}}$nrich R., 2016, MNRAS, 462, 1697

\noindent Bahcall J.N., 1984a, ApJ, 276, 156

\noindent Bahcall J.N., 1984b, ApJ, 276, 169

\noindent Bahcall J.N., 1984c, ApJ, 287, 926

\noindent Bahcall J.N., Flynn C., Gould A., 1992, ApJ, 389, 234

\noindent Banerjee A., Jog C.J., 2007, ApJ, 662, 335

\noindent Barbanis B., Woltjer L., 1967, ApJ, 150, 461

\noindent Bienaym$\mathrm{\acute{e}}$ O. et al., 2014, A\&A, 571, A92 

\noindent Binney J., Tremaine S., 1987, Galactic Dynamics, Princeton Univ. Press,
Princeton, NJ

\noindent Binney J. et al., 2014, MNRAS, 439, 1231

\noindent Bond N. A. et al., 2010, ApJ, 716, 1 

\noindent Bovy J., 2017, MNRAS, 470, 1360

\noindent Camm G.L. 1950, MNRAS, 110, 305

\noindent Carlberg R. G., Sellwood J. A. 1985, ApJ, 292, 79

\noindent Clemens D. P., 1985, ApJ, 295, 422

\noindent Comer$\mathrm{\acute{o}}$n S et al., 2011, ApJ, 741, 28

\noindent Cr$\mathrm{\acute{e}}$z$\mathrm{\acute{e}}$ M., Chereul E., Bienaym$\mathrm{\acute{e}}$ O., Pichon C., 1998, A\&A, 329, 920

\noindent Dehnen W., Binney J., 1998, MNRAS, 298, 387

\noindent Ferguson D., Gardner S., Yanny B., 2017, ApJ, 843, 141

\noindent Fuchs B. et al., 2009, AJ, 137, 4149 

\noindent Gaia collaboration et al., 2018, A\&A 616, A11

\noindent Garbari S., Read J.I,, Lake G., 2011, MNRAS, 416, 2318

\noindent Gilmore G., Reid N., 1983, MNRAS, 202, 1025

\noindent Guo R. et al., 2020, MNRAS, 495, 4828

\noindent Gustafsson B., Church R.P., Davies M.B., Rickman H., 2016, A\&A, 593, A85

\noindent Hagen J.H.J,  Helmi A., 2018, A\&A, 615, A99

\noindent Holmberg J., Flynn C., 2000, MNRAS, 313, 209

\noindent Jenkins, A., Binney, J. 1990, MNRAS, 245, 305

\noindent Jing Y. et al., 2016, MNRAS, 463, 3390

\noindent Kalberla P.M.W, 2003, ApJ, 588, 805

\noindent Saha K., Tseng Y-H., Taam R.E., 2010, ApJ, 721, 1878

\noindent Kuijken K., Gilmore G., 1989, MNRAS, 239, 651

\noindent Lacey C.G., 1984, MNRAS, 208, 687

\noindent Lewis B. M., 1984, ApJ, 285, L453

\noindent Lewis J. R., Freeman K. C., 1989, AJ, 97, 139

\noindent Matthews L. D., 2000, AJ, 120, 1764

\noindent Ma X et al., 2017, MNRAS, 467, 2430

\noindent Mera D., Chabrier G., Schaeffer R., 1998, A\&A, 330, 953

\noindent Mignard F., 2000, A\&A, 354, 522

\noindent Narayan C. A., Jog C. J., 2002, A\&A, 394, 89

\noindent Oort J.H., 1960, B.A.N., 15, 45

\noindent Perry C.L., 1969, AJ, 74, 139

\noindent Rohlfs K., 1977, Lectures on Density Wave Theory (Berlin: Springer-Verlag)

\noindent Salomon J.-B., Bienaym$\mathrm{\acute{e}}$ O., Reyl$\mathrm{\acute{e}}$ C., Robin A.C., Famaey B., 2020, 
preprint (arXiv:2009.04495v1)

\noindent Sarkar S., Jog C. J., 2018, A\&A, 617, A142

\noindent Sarkar S., Jog C.J., 2019, A\&A, 628, A58

\noindent Sarkar S., Jog C.J., 2020, MNRAS, 492, 628

\noindent Scoville, N. Z.,  Sanders, D. B. 1987, in Interstellar Processes, eds. D. J.
Hollenbach, \& H. A. Thronson (Dordrecht: Riedel), 21

\noindent Sharma S. et al., 2020, preprint (arXiv:2004.06556v1)

\noindent Sun W.-X. et al., 2020, preprint (arXiv:2008.10218v1)

\noindent Spitzer L., 1942, ApJ, 95, 329

\noindent Spitzer, L. 1978, Physical Processes in the Interstellar Medium (New York: John Wiley)

\noindent Stark, A. A. 1984, ApJ, 281, 624

\noindent van der Kruit, P. C., 1988, A\&A, 192, 117

\noindent Vel$\mathrm{\acute{a}}$zquez H., White S.D.M., 1999, MNRAS, 304, 254

\noindent Walker I.R., Mihos J.C., Hernquist L., 1996, ApJ, 460, 121

\noindent Wang, H-F et al, 2018, MNRAS, 478, 3367

\noindent Wielen R., 1977, A\&A, 60, 263

\noindent Wu J., Struck C., D'onghia E., Elmegreen B.G, 2020, preprint (arXiv:2009.01929v1)

\noindent Xia Q et al., 2016, MNRAS, 458, 3839

\noindent Yoachim P., Dalcanton J.J., 2006, AJ, 131, 226

\bigskip

\appendix 
\section{Effect of non-isothermal vertical velocity dispersion on the asymmetric drift}
\label{sec:Appendix A}

The asymmetric drift or the apparent lag of the rotational velocity w.r.t the true circular velocity of a stellar population, is defined as $v_{a}\equiv v_{c}-\overline{v_{\phi}}$, where $v_{c}$ is the true circular speed at a radius $R$ and $\overline{v_{\phi}}$ is the mean azimuthal velocity/ rotation velocity of the stellar population, in the cylindrical co-ordinate system. Now using the axisymmetric radial Jeans equation one can calculate the asymmetric drift (Binney \& Tremaine 1987), evaluated at $z$=0, as 

\begin{equation}
v_{a} \simeq \frac{\overline{v^{2}_{R}}}{2v_{c}}\left[\frac{\sigma^{2}_{\phi}}{\overline{v^{2}_{R}}}-1-\frac{\partial \mathrm{ln}\left(\nu(R,z) \overline{v^{2}_{R}}\right)}{\partial \mathrm{ln} R} - \frac{R}{\overline{v^{2}_{R}}}\frac{\partial}{\partial z}(\overline{v_{R}v_{z}})  \right] \label{eq:A1}
\end{equation}

\noindent where the disc is assumed to be in a steady state and symmetric about $z$=0. 

The mean radial velocity is denoted by $\overline{v_{R}}$ which satisfies $\sigma^{2}_{R}=\overline{v^{2}_{R}}-\overline{v_{R}}^{2}$. Similarly, $\sigma^{2}_{z}=\overline{v^{2}_{z}}-\overline{v_{z}}^{2}$ and $\sigma^{2}_{\phi}=\overline{v^{2}_{\phi}}-\overline{v_{\phi}}^{2}$, here $\sigma_{i}$ denotes the velocity dispersion along the three axes. The number density distribution is denoted by $\nu(R,z)$. The cross term of the velocity ellipsoid, i.e, $\overline{v_{R}v_{z}}$ is given as $(\overline{v^{2}_{R}}-\overline{v^{2}_{z}})z/R$ (Binney \& Tremaine 1987) when the velocity ellipsoid is tilted w.r.t to the R-z axes and is aligned with a spherical system centered at the center of the Galaxy (as observed in most of the recent data). 
For a cylindrical alignment the last term in Eq.(A1) vanishes (Binney \& Tremaine 1987). We assume the mean motions along $R$ and $z$ to be zero and therefore obtain $\sigma^{2}_{R}=\overline{v^{2}_{R}}$, $\sigma^{2}_{z}=\overline{v^{2}_{z}}$ and $\overline{v_{R}v_{z}}=(\sigma^{2}_{R}-\sigma^{2}_{z})z/R$, this last expression is obtained considering the axes of the velocity ellipsoid to be aligned along the axes of the spherical co-ordinates
centered on the galactic center. Putting these in the Eq.(A1), we obtain 

\begin{equation}
v_{a} = \frac{\sigma^{2}_{R}}{2v_{c}}\left[\frac{\sigma^{2}_{\phi}}{\sigma^{2}_{R}}-1-\frac{\partial \mathrm{ln}\left(\nu(R,z) \sigma^{2}_{R}\right)}{\partial \mathrm{ln} R} - \frac{R}{\sigma^{2}_{R}}\frac{\partial}{\partial z}\left(\frac{(\sigma^{2}_{R}-\sigma^{2}_{z})z}{R}\right)  \right]. \label{eq:A2}
\end{equation}

\noindent Since all the terms are evaluated at $z$=0, it can be easily seen that the first three terms are independent of the vertical variation of the dispersions.

To evaluate the last term at $z$=0, first we need to calculate the derivative as follows

\begin{equation}
\frac{R}{\sigma^{2}_{R}}\frac{\partial}{\partial z}\left(\frac{(\sigma^{2}_{R}-\sigma^{2}_{z})z}{R}\right) = \left(1- \frac{\sigma^{2}_{z}}{\sigma^{2}_{R}}\right)+\frac{z}{\sigma^{2}_{R}}\frac{\partial}{\partial z}\left(\sigma^{2}_{R}-\sigma^{2}_{z}\right).\label{eq:A3}
\end{equation}

\noindent Here the first term depends on the dispersion values at the mid-plane only. The second term is clearly zero for isothermal dispersions. For non-isothermal vertical dispersion we have considered $\sigma_{z}=\sigma_{z,0}+Cz$ in our model (Eq.(7) in Section 2.1). Similarly, we assume $\sigma_{R}=\sigma_{R,0}+C_{1}z$. Therefore the second term can be calculated as

\begin{equation}
\frac{z}{\sigma^{2}_{R}}\frac{\partial}{\partial z}\left(\sigma^{2}_{R}-\sigma^{2}_{z}\right)=\frac{z}{\sigma^{2}_{R}}\left[2C_{1}\left(\sigma_{R,0}+C_{1}z\right)-2C\left(\sigma_{z,0}+Cz\right)\right].\label{eq:A4}
\end{equation}  

\noindent The above expression is thus zero at $z$=0.

Thus, the asymmetric drift is found to remain independent of the variation of the dispersions along the $z$ direction, that is, it is not affected by the non-isothermal velocity dispersion. We have shown it for the linearly increasing case as considered in our model, and this conclusion will be valid for other usually observed forms where the dispersion increases with $z$.

\end{document}